\documentclass[sigconf]{acmart}
 
\usepackage{booktabs} % For formal tables
\usepackage{url}  %Required
\usepackage{graphicx}  %Required
\usepackage{color}
\usepackage{hyperref}
\usepackage{tikz}
\usepackage{pgfplots}
\usepackage{float}
\usepackage{amsmath,amssymb}
\usepackage{color}
\usepackage{booktabs}
\usepackage{multirow}
\usepackage{bbm}
\usepackage[linesnumbered,ruled]{algorithm2e}
\usepackage{caption}
\usepackage{subfig}

\setcopyright{rightsretained}
\newcommand\T{\rule{0pt}{2.9ex}}       % Top strut
\newcommand\B{\rule[-1.2ex]{0pt}{0pt}} % Bottom strut

\acmDOI{}

\acmISBN{}

\acmConference[RecSys'18]{ACM Conference on Recommender Systems}{October 2018}{Vancouver, Canada}
\acmYear{2018}
\copyrightyear{2018}

\acmArticle{4}
\acmPrice{15.00}

\begin{document}
\title[RecoGym]{RecoGym: A Reinforcement Learning Environment for the Problem of Product Recommendation in Online Advertising}

\author{David Rohde}
\orcid{1234-5678-9012}
\affiliation{%
  \institution{Criteo AI Labs}
  \city{Paris}
}
\email{d.rohde@criteo.com}

\author{Stephen Bonner}
\authornote{Author is doctoral student at The Department of Computer Science, Durham University, UK}
\orcid{1234-5678-9012}
\affiliation{%
  \institution{Criteo AI Labs}
  \city{Paris}
}
\email{st.bonner@criteo.com}

\author{Travis Dunlop}
\authornote{Work done will interning at Criteo AI Labs.} 
\affiliation{%
\institution{Universitat Pompeu Fabra}
  \city{Barcelona}
}
 \email{dunloptravis@gmail.com}
 
\author{Flavian Vasile}
\affiliation{%
\institution{Criteo AI Labs}
  \city{Paris}
}
 \email{f.vasile@criteo.com}

 \author{Alexandros Karatzoglou}
 \affiliation{%
 \institution{Telefonica Research}
   \city{Barcelona}
 }
  \email{alexandros.karatzoglou@gmail.com}

% \author{Stephen Bonner}
% \authornote{Author is doctoral student at The Department of Computer Science, Durham University, UK}
% \affiliation{%
%   \institution{Criteo Research}
%   \city{Paris}
% }
% \email{st.bonner@criteo.com}

% \author{Flavian Vasile}
% \affiliation{%
%   \institution{Criteo Research}
%   \city{Paris}
% }
% \email{f.vasile@criteo.com}

%\renewcommand{\shortauthors}{S. Bonner \& F. Vasile}

\begin{abstract}
Recommender Systems are becoming ubiquitous in many settings and take
many forms, from product recommendation in e-commerce stores, to query
suggestions  in search engines, to friend recommendation in social
networks. Current research directions which are largely based upon supervised learning from
historical data appear to be showing diminishing returns, with many practitioners reporting a discrepancy between improvements in offline
metrics for supervised learning and the online performance of the
newly proposed models. One possible reason is that we are using the wrong paradigm: when looking at the long-term cycle of collecting historical performance data, creating a new version of the recommendation model, A/B testing it, and then rolling it out, we see
that there a lot of commonalities with the reinforcement learning (RL)
setup, where the agent observes the environment and acts upon it in
order to change its state towards better states (states with higher
rewards).  To this end  we introduce \emph{RecoGym}, an RL environment
for recommendation, which is defined by a model of user traffic
patterns on e-commerce and the users response
to recommendations on the publisher websites. We believe that this is
an important step forward for the field of recommendation systems
research, that could open up an avenue of collaboration between the
recommender systems and reinforcement learning communities and lead to
better alignment between offline and online performance metrics. 
\end{abstract}

\maketitle

%----------------------------------------------------------------------------------------
%	SECTION - Introduction
%----------------------------------------------------------------------------------------
\section{Introduction}
Recommender systems are becoming more and more prevalent in all walks of modern life. They are particularly valuable when a user would otherwise be forced to do exhaustive searches in spaces with huge numbers of available options (either in terms of knowing what are the best products to buy, people to befriend, jobs to apply for) and that ultimately leads to user unhappiness (see \emph{The Paradox of Choice} \cite{schwartz2004paradox}).  The theory of learning optimal recommendation policies is also evolving quite rapidly, and the limitations of the classical supervised approach to learning what to recommend are becoming apparent. Practitioners are increasingly reporting poor correlations between offline metrics and measured online performance. 

In the context of online performance advertising, recommendation appears as the problem of choosing what item to advertise to the user given her current state and given the current context in which the ad is presented (display medium, geography, time of day, etc.). By the nature of the problem, the user has two types of interactions with the items: first, the interactions on the e-commerce website, which we denote as \emph{organic sessions} and secondly, the interactions on the publisher websites, through ads, which we denote as \emph{bandit sessions}, inspired by the bandits literature which is represented heavily in the computational advertising field.

The problem of learning what product/item ads to show to each user is addressed differently in the recommendation literature and computational advertising literature. In the recommendation literature, this would be either solved as a \emph{missing link prediction task} or as a \emph{next item prediction task}, if time is important. The associated metrics for these approaches are classification/regression metrics such as area under the curve, mean squared error and negative log likelihood and respectively, ranking metrics such as precision@k and negative discounted cumulative gain. 

Modern approaches to computational advertising employ randomized policies that ensure that exploration of actions using strategies such as epsilon-greedy or Thompson sampling. By adding randomization to the recommendation policy, we are able to not only compute the probability of an ad under the current policy, but also under any other recommendation policy using \emph{Inverse Propensity Scoring} (IPS) \cite{rosenblum1983central}. This type of counterfactual estimation is explored in a series of papers \cite{li2010contextual}, \cite{dudik2011efficient} \cite{joachims2-18deep} and to date, one item recommendation dataset with inverse propensity scores has been released, see \cite{lefortier2016large}.

However, these approaches suffer from problems when the recommendation policy to be evaluated is far from the logging policy, which results in very large inverse propensity scores for some of the items and due to capping, to bias in the estimation of the performance of the new policy.

To address the issues of IPS-based and other classical offline evaluation approaches, we propose \emph{RecoGym}\footnote{\url{https://github.com/criteo-research/reco-gym}}, a full reinforcement learning (RL) environment simulator for recommendation that allows, at evaluation time, the simulation of users' reaction to arbitrary recommendation policies, without incurring the issue of exploding variance that is present in the case of datasets with IPS weights.

Since we are releasing this as an OpenAI Gym environment \cite{brockman2016openai}, we hope that this will allow both recommendation and RL practitioners to expand the type of machine learning approaches for recommendation and bring RL to other real-world applications, outside of games and self-driving cars. 

In the first version of our simulator, we are releasing an environment that has the following key properties:
\begin{itemize}
\item The environment takes into account both the organic and the bandit user-item interactions.
\item It contains a parameterized level of correlation between the two types of user behavior.
\item It allows to parameterize the dimension of the hidden space in which the users and items are clustered.
\item It allows to parameterize the level of relative impact of the user's past exposure level to ads on the click-through rate of an ad display at a given time
\end{itemize}

The rest of the paper is organized as follows: in Section \ref{relwork} we review related work; in Section \ref{game} we introduce the reinforcement learning problem.  In Section \ref{simulator} we provide some basic sanity checks that we expect a successful model of user reaction to in-ad recommendations to pass, and in Section \ref{conc} we conclude with an overview of contributions and planned next steps for future releases.

%----------------------------------------------------------------------------------------
%	SECTION - Introduction
%----------------------------------------------------------------------------------------

\section{Related Work}
\label{relwork}
The relevant literature is currently split into two distinct parts: the classical recommendation literature, that considers modeling the \emph{organic user behavior} and the computational advertising literature that models the \emph{bandit user behavior}.  

The recommendation literature on modeling organic user behavior is vast: often, organic recommendation is framed as a user-item matrix completion task with matrix factorization as a common approach \cite{mf, prod2vec}.  Time-aware organic models use sequential models such as recurrent neural networks in order to predict next item viewed; see \cite{quadrana2018sequence} for a recent survey.

Some works focused on organic datasets are aware of the tension between the task of modeling organic user behavior and the final goal of developing a ranking over potential recommendations.  The ranking of organic user behavior is often used as a proxy to ranking bandit behavior - for a prominent example see \cite{bpr}.  In a similar spirit, \cite{Bonner2017CausalEF} and \cite{liangcausal} use the fact that the scoring phase of a recommendation engine must uniformly evaluate every action in order to adjust the organic model to be more appropriate for making recommendations.  There are numerous public datasets of organic user behavior e.g. \cite{harper2016movielens} \cite{bennett2007netflix},  including some with temporal information\cite{quadrana2018sequence}.  

The recommendation literature on modeling bandit user behavior is smaller: contextual bandits is a popular formulation of this type of problem
\cite{li2010contextual}, \cite{dudik2011efficient}; as is counter factual risk minimization \cite{joachims2-18deep}. Another approach is Gaussian process bandit optimization \cite{krause2011contextual}, \cite{kawale2015efficient}.  There are also relatively few datasets for benchmarking such algorithms. \cite{lefortier2016large} is one exception, unfortunately the requirement to use IPS for evaluation means that the measures of algorithmic performance are noisy.

Very importantly, to the best of our knowledge, there are currently no public datasets or indeed no other way to evaluate models that is able to incorporate both organic and bandit information.

\section{Formalizing Recommendation as a Game}
\label{game}
\begin{figure}[h!]
    \includegraphics[scale=0.25]{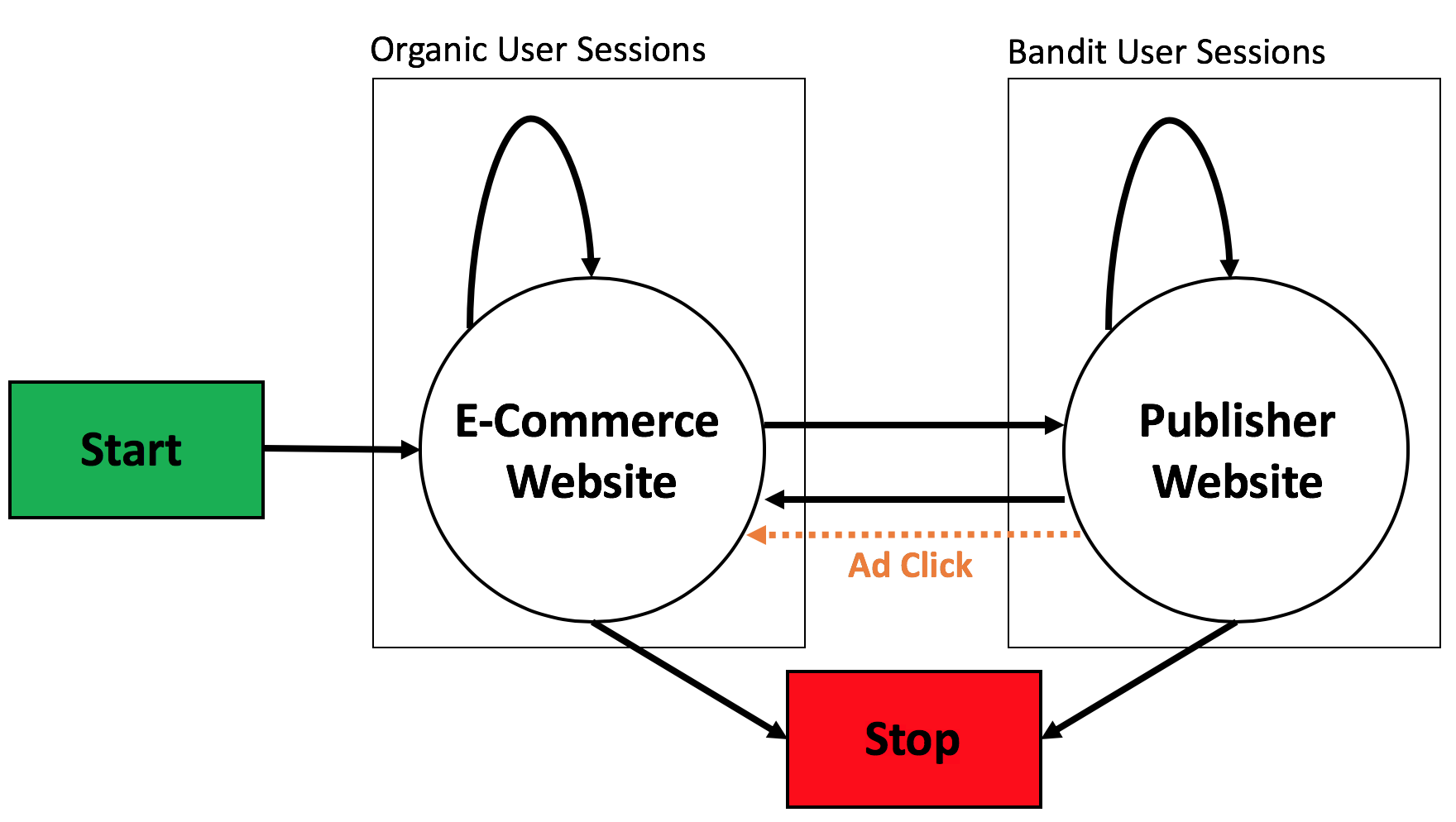}
      \centering
      \caption{Markov Chain of the organic and bandit user sessions}
      \label{fig:influence}
   \end{figure}

In Figure \ref{fig:influence} we show our idealized (and heavily simplified) model of user activity, that alternates between organic e-commerce sessions and bandit publisher sessions and finishes by stopping her online activity for a longer period of time. In this setup, the advertising recommendations can only be shown during the publisher sessions. For simplicity, in this version of the model we assume that the user has a constant probability of conversion once she is back on one of the product shopping pages.  

Under this formulation of the environment and attached rewards, the goal of the recommendation agent is to show personalized ads that increase the likelihood of the user to transition back to the e-commerce website, e.g. to increase the click-through rate of the user-ad pairs  (shown as a dotted arrow in the figure below). 

  %\begin{figure}[h!]
  %  \includegraphics[scale=0.25]{Figures/organic_diag.png}
  %    \centering
  %    \caption{FSM of organic user behavior}
  %    \label{fig:control}
  %\end{figure}
  
   % \begin{figure}[h!]
   %\includegraphics[scale=0.25]{Figures/influence_diag.png}
   %   \centering
   %   \caption{FSM of ad-influenced user behavior}
   %   \label{fig:influence}
   %\end{figure}
    
\subsection{Notation}

\begin{table}
\begin{tabular}{ c p{6.5cm} }
\toprule
\textbf{Symbol} & \textbf{Meaning} \\
\midrule \midrule
$P$ & Number of products\\
$u$ & User \T \\
$t$ & Time index (discrete) starting from 0\\
$z_{u,t}$ & Event $t$ for user $u$ is organic or bandit\\
$v_{u,t}$ & Organic product view or undefined if $z_{u,t}={\rm bandit}$\\
$v_{u,p,t}$ & One hot coding of $v_{u,t}$ where $p$ denotes the product\\
$a_{u,t}$ & Recommended or action product or undefined if $z_{u,t}={\rm organic}$\\
$c_{u,t}t$ & Click was made on recommendation $a_{u,t}$ or\\
& undefined if $z_{u,t}={\rm organic}$ \B \\

\bottomrule
\end{tabular}
\caption{Notation}
\label{notation}
\end{table}

\begin{table}
\begin{tabular}{ l l l l l l}
\toprule
$u$ & $t$ & $z_{u,t}$ & $v_{u,t}$ & $a_{u,t}$ & $c_{u,t}$\\
\midrule \midrule
. & . & . & . & . &. \T \\
10 & 0 & organic & 104 & NA & NA \\
10 & 1 & organic & 52 & NA & NA \\
10 & 2 & organic & 71 & NA & NA \\
10 & 3 & bandit & NA & 42 & 0 \\
10 & 4 & bandit & NA & 52 & 1 \\
10 & 5 & organic & 52 & NA & NA \\
. & . & . & . & . &. \B \\

\bottomrule
\end{tabular}
\caption{Example Data}
\label{example}
\end{table}

A table of notation used is given in Table \ref{notation}.  An example of the type of data is given in Table \ref{example}.  

In the example we see the timeline of user $10$ over 6 time steps.  All users start with an organic event at time 0 and in this case, user 10 visits product 104, then she transitions at time step 1 to product 52 and at time step 2 to product 71.  At this point the user stops browsing the retailer website and her organic session ends; she does however go to a publisher website and we target her with ads starting with a recommendation/action($a$) for product 42 at time step 3, which she does not click on.  Following this unsuccessful recommendation, another recommendation is made at time step 4 for product 52 and in this case the user responds positively and clicks.  The clicking event brings the user back to the retailer website and her organic session restarts at time step 5 with an organic view for product 52.
 
%----------------------------------------------------------------------------------------
%	SECTION - Simulator
%----------------------------------------------------------------------------------------
\section{Simulator}
\label{simulator}
\subsection{Reinforcement Learning Environment}

The modern advances in combining deep neural networks with RL has been enabled by the ability of various algorithms to have access to massive quantities of simulated data from which to learn. For example, the recent success of Deep Mind's AlphaGo was driven by the ability of the algorithm to compete, and crucially, learn from many millions of simulated games of Go \cite{silver2017mastering}. As such, simulated environments in which RL algorithms can learn from a near unlimited data source are becoming key to the creation of state-of-the-art algorithms. However, to date, there has been little work on designing an RL environment specifically for the training and evaluation of recommender systems. Our work will focus on designing and implementing such an environment.

In 2016, OpenAI launched the OpenAI Gym, a tool-kit designed to create a unified API in which many RL benchmark problems can be implemented, allowing for direct comparisons between RL approaches \cite{brockman2016openai}. OpenAI Gym draws upon earlier work on providing RL benchmarks, for example the Arcade Learning Environment where two dimensional video games were used as problems for an agent to solve \cite{bellemare2013arcade}. Currently OpenAI Gym provides pre-made gym environments in areas such as control theory and simulated robotics. However, users are able to implement their own gym environments for any arbitrary task via the unified Python-based gym API. 

We propose a gym-based environment specifically for the task of computational advertising focused recommender systems. To create an OpenAI Gym compliant environment, one must implement several standard gym functions including Step and Reset \cite{brockman2016openai}. The gym API assumes that an \emph{Agent} class has been implemented by the end-user to allow their chosen RL algorithm to interface with the gym environments. The primary tasks of a gym environment are to produce observations given the current state and to ingest actions from the agents to produce reward signals. We will map the required OpenAI Gym functions for a \emph{Environment} class to our recommendation simulator in the following manner:  

\begin{itemize}
    \item \emph{Reset} - When called, the simulator initializes a random synthetic user. 

    \item \emph{Step} - At each step call, the simulator ingests the predicted product recommendation from the agent and, as specified in the Open AI Gym white paper, returns four objects \cite{brockman2016openai}: \begin{itemize} 
        
        \item \emph{Observation} - This returns the users complete last organic session if available. When bandit events are occurring this will return \emph{None}. 
    
        \item \emph{Reward} - The reward given by the environment for the previous recommendation i.e. did the recommendation result in a click?

        \item \emph{Done} - A Boolean value denoting if the user's shopping sequence has finished. If true, the \emph{Reset} function will be called and a new user will be generated for the next step.  

        \item \emph{Info} - Potential logging information from the simulation. \end{itemize}

    Following a call to reset, step will take no action and return the first batch of organic user behavior. In subsequent calls, the action and the reward will be defined.
\end{itemize}

\subsection{Sanity Checks for an Agent that Combines Organic and Bandit information}

The most successful algorithm will combine organic and bandit
information in order to make good recommendations and obtain a low
regret.  Basic sanity checks for such an algorithm include that in the
limit of large amounts of data it performs similarly to a pure bandit
algorithm and that in the presence of less bandit data it is able to
gain some advantages from the organic data.

In order to formalize these properties we describe three models; one that uses only the organic data, one that
uses the bandit and a third that uses both attempts to get the best of both worlds.

\begin{figure}%
    \centering
    \subfloat[Performance as \#bandit events increases]{{\includegraphics[width=8cm]{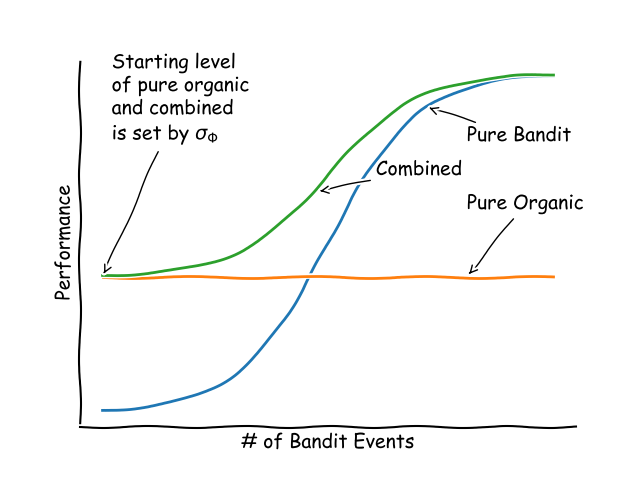} }}%
    \qquad
    \subfloat[Performance as $\sigma_\Phi$ increases]{{\includegraphics[width=8cm]{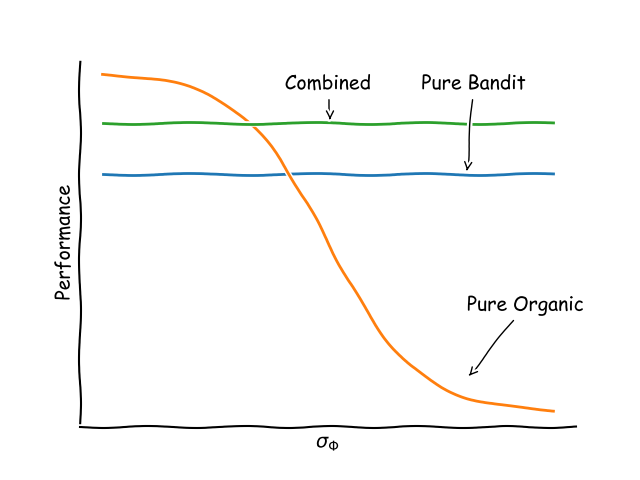} }}%
    \caption{Sketch of expected performance}%
    \label{fig:bandit_organic}%
\end{figure}

Let $v_{u,p,t} =1$ if user $u$ organically viewed
product $p$ at time $t$ and $v_{u,p,t} =0$ otherwise.  Furthermore
define $\Lambda_{u,p,t}$ such 
that $v_{u,p,t} \sim {\rm Bernoulli}(\sigma( \Lambda_{u,p,t} ))$.
Here $\sigma(\cdot)$ is the logistic sigmoid and  $\Lambda_{u,p,t}$ is
the log odds of user $u$ organically viewing product $p$ at time $t$.
This allows us to draw a formal link between the organic world and the
bandit world in the equation below:

\[
\Phi_{u,a,t} = f(\Lambda_{u,p,t} + \epsilon_{u,a,t}),
\]

\noindent
where $\Phi_{u,a,t} $ is the click through rate for giving
action/recommendation $a$ to user $u$ at time $t$;
$\epsilon_{u,a,t}$ is ``noise'' with mean zero and variance
$\sigma_{\Phi}^2$.   The
function $f(\cdot)$ is an increasing function that calibrates organic behavior
(which has a categorical distribution) to bandit behavior which has a
Bernoulli distribution.  The fact that $\Phi_{u,a,t}$ is indexed by
time $t$ means that non-stationary effects such as ad-fatigue i.e. the
user loosing interest in an ad after repeated displays can be incorporated.

When $\sigma_{\Phi}$ is zero, so is
$\epsilon_{u,a,t}$ and in order to make a recommendation
we simply need to compute $a^* = {\rm argmax}_p  \hat{\Lambda}_{u,p,t}$, where
$\hat{\Lambda}$ can be estimated from purely organic data.  While
organic data alone is insufficient for estimating $\Phi$ due to the
fact that the form of $f(\cdot)$ is unknown it {\emph is sufficient}
for correctly ordering recommendations.  It is
precisely  in this case where purely organic algorithms are able to
perform well, and organic metrics such as precision@k in should be
expected to correlate with click through rate. 
We call these sorts of algorithms ``Pure Organic'' algorithms.

In contrast when $\sigma_{\Phi}$ increases organic data alone becomes
insufficient in order to determine the best actions as $\Phi_{u,1,t},
..., \Phi_{u,P,t}$ will typically have a different ordering to
$\Lambda_{u,1,t}... \Lambda_{u,P,t}$.

An extreme approach in this situation is to use a ``Pure Bandit''
algorithm that estimates $\Phi$ directly and ignores the decomposition
into $\Lambda$ and $\epsilon$.  This approach should be expected to
work when bandit data is plentiful, but ignores the advantages that
combining the two datasets together may provide.  These advantages
would be realized by a ``Combined'' algorithm that may use two main
principles in order to infer best actions more quickly than ``Pure
Bandit'' and ``Pure Organic'':

\begin{enumerate}
\item
If $\sigma_{\Phi}$ is not too high the ordering provided by
$\hat{\Lambda}$ may be approximately correct.  Bandit events then can
be viewed as refining the ordering of these recommendations.
\item
We may find (or assume) that the correlation structures in $\epsilon$
are similar to those in $\Lambda$, i.e. if users or products are
observed to be correlated organically then we may (subject to
confirmation in the data) assume that a similar correlation exists
between users and actions in $\epsilon_{u,a,t}$.  In the first version
of the simulator this will be implemented by users belonging to hidden latent variables or clusters.
\end{enumerate}

\noindent
Now that we have formalized the connection between bandit and organic
we can sketch the relationships we expect to observe as we vary (a)
the amount of bandit data available for training and (b) the amount of
noise $\sigma_{\Phi}$.  The relationships we expect to see are shown
in Figure \ref{fig:bandit_organic}.  

In Figure
\ref{fig:bandit_organic} a) we see that the ``Pure Organic'' behavior
is determined by the amount of noise in $\sigma_{\Phi}$ and is
unaffected by the number of bandit events.   If $\sigma_{\Phi}$ is
sufficiently small that the actions are correctly ordered it will
perform well, in contrast if it is large it will perform poorly.  The
performance does not change as the number of bandit events increase as
these are not used.  In contrast the ``Pure Bandit'' algorithm is
unable to make predictions without large numbers of bandit events.  It
performs poorly when this data is scarce and well when it is
plentiful.  The ``Combined'' model should achieve a happy medium
using information from both sources and therefore able to outperform
``Pure Bandit'' for moderate numbers of bandit events the two agreeing
only in the limit of (often) unfeasibly large numbers of bandit
events.

In Figure
\ref{fig:bandit_organic} a) we see that the ``Pure Organic''
performance reduces as $\sigma_{\Phi}$ breaks the connection between
organic and bandit events.  In contrast the combined and pure bandit
do not exhibit this issue.  As ``Pure Bandit'' ignores organic
information we expect it to be invariant to $\sigma_{\Phi}$ and its
performance will be determined by the amount of bandit events.  Again
we expect a ``Combined'' model to exhibit better performance so long
as bandit events are not too plentiful.

\subsection{Baseline Agents}

In addition to the reco-gym environment, we also provide a selection of baseline agents to interact with it. These agents include the following:

\begin{itemize}

    \item \emph{Random} - This agent makes random recommendations and performs no learning from the organic data.

    \item \emph{Logistic} -  During training, this agent maintains a count of how many times a given user is exposed to a certain recommendation, as well as their response (click/no-click). Here the user is represented by their last organically viewed product, ignoring any information from the users organic sequence. To recommend a product, the logistic agent simply looks up the product with the highest CTR for the current user. 

    \item \emph{Supervised-Prod2Vec} - Inspired by the prod2vec algorithm \cite{grbovic2015commerce}, we implement a supervised user-product factorization like agent. In this agent, a user is again represented by their last organically seen product only. Clicks are predicted as a sigmoid over a linear transform of the user representation (their last organic product seen) and the recommended product representation.

\end{itemize}
\section{Conclusions}
\label{conc}

In this paper we introduce \emph{RecoGym}, the first Reinforcement Learning environment for recommendation in the context of online advertising. In the first release, the simulator supports the existence of user shopping and user publisher sequences, it allows the experimenter to specify the desired level of correlation between the two, to simulate various intrinsic dimensions for the user - item clusters and to vary the impact of repeated ad exposure on the ad click-through rate. In the paper we additionally cover some basic sanity checks on the behavior of the models trained on the simulator data and propose guidelines. Finally, we hope the release of this simulator will bring more interdisciplinary research between the reinforcement learning and recommender system communities and that subsequently, better alignment between offline and online performance can be achieved. 

\bibliographystyle{ACM-Reference-Format}
\bibliography{literature} 

\end{document}